\documentclass[11pt,twoside]{article}
\usepackage{rjr-asp}
\usepackage{lscape}
\usepackage{epsfig,graphicx,natbib,url}  
\begin{document}

\markboth{Ichimoto et al.}{On-orbit performance of SOT}
\title{On-orbit Performance of the Solar Optical Telescope aboard Hinode}  
\author{K.Ichimoto$^{1}$, Y.Katsukawa$^1$, T.Tarbell$^2$, R.A.Shine$^2$,
        C.Hoffmann$^2$, T.Berger$^2$, T.Cruz$^2$, Y.Suematsu$^1$, S.Tsuneta$^1$, 
	T.Shimizu$^3$, B.W.Lites$^4$}

\affil{$^1$ National Astronomical Observatory of Japan, 
	2-21-1, Osawa, Mitaka, Toyko, 181-8588, Japan \\
       $^2$ Lockheed Martin Advanced Technology Center, 
	3251 Hanover Street, Palo Alto, CA 94304, USA \\
       $^3$ Japan Aerospace Exploration Agency, 
	Institute of Space and Astronoutical Science, 
	3-1-1, Yoshinodai, Sagamihara, Kanagawa, 229-8510, Japan\\
       $^4$ High Altitude Observatory, National Center for Atmospheric Research, 
	P.O. Box 3000 Boulder, CO 80307-3000, U.S.A \\
	}
  
\begin{abstract} 
On-orbit performance of the Solar Optical Telescope (SOT) aboard Hinode 
is described with some attentions on its unpredicted aspects. 
In general, SOT revealed an excellent performance and has been providing outstanding data.
Some unexpected features exist, however, in behaviors of the focus position, 
throughput and structural stability.
Most of them are recovered by the daily operation 
i.e., frequent focus adjustment, careful heater setting and corrections in data analysis.
The tunable filter contains air bubbles which degrade the data quality significantly.
Schemes for tuning the filter without disturbing the bubbles 
have been developed and tested, and some useful procedures to obtain Dopplergram
and magnetogram are now available.
October and March when the orbit of satellite becomes nearly perpendicular 
to the direction towards the sun provide a favorable condition 
for continuous runs of the narrow-band filter imager.
\end{abstract}

\section{Introduction}      \label{a}

The Solar Optical Telescope (SOT, Tsuneta et al. 2007) 
onboard Hinode (Kosugi et al 2007)
aims to provide high resolution images and high precesion 
magnetograms of the solar photosphere
and chromosphere under the unprecedentedly stable conditions from space.
The SOT consists of the Optical Telescope Assembly (OTA, Suematsu et al. 2007) 
and the Focal Plane Package (FPP, Tarbell et al. 2007).
Four optical paths are in the FPP; Broad-band Filter Imager (BFI), 
Narrow-band Filter Imager (NFI), Spectro-polarimeter(SP), and Correlation Tracker (CT).
Since 25th Oct. 2006, the day of the successful first light, the SOT has been
sending remarkable data of the sun.
The excellent quality of the data provided by the SOT is reported by
Suematsu etal (2007) and Shimizu etal (2007) and also is evident from 
the outstanding results of the on-going researches presented in this conference.
In this paper, we report the on-orbit performance of the SOT with some
emphasis on unpredicted features of the instrument.

\section{On-orbit performance of the SOT}     \label{ichimoto-sec:image}

\subsection{Image quality and pointing stability} 

Figure 1 shows a G-band (430nm) image taken by the BFI.
The magnified image on the right shows a bright point in between granulations
as the smallest feature in the field of view.
The size of this structure is fairly close to that of the theoretical point spread
function and we conclude that the SOT achieved 'diffraction limited' optical
performance (Strehl ratio $>$0.8, Suematsu et al 2007).
More quantitative evaluations using the phase diversity method are on-going.

Figure 2 shows time profiles of the CT signal, i.e., the image displacement
with respect to a fixed reference image taken by the CT camera at 580Hz 
(Shimizu et al 2007).
During the period when the CT-servo is on, the image stability gets
as high as 0.01arcsec(rms) in both X- and Y-directions, which is 
about three times better than the requirement.
Moving mechanisms in the three telescopes do not produce a significant 
disturbance on the SOT image during their movement except the 
visible-light shutter (VLS) of the XRT, 
which produces a SOT image jitter of about 0.4 arcsec rms during 
the period of its movement ($\sim$0.5 sec).
The impact of the XRT-VLS on the SOT observation is negligible
since the frequency of its usage is sufficiently low.

\begin{figure}  
  \centering
  \includegraphics[width=9cm]{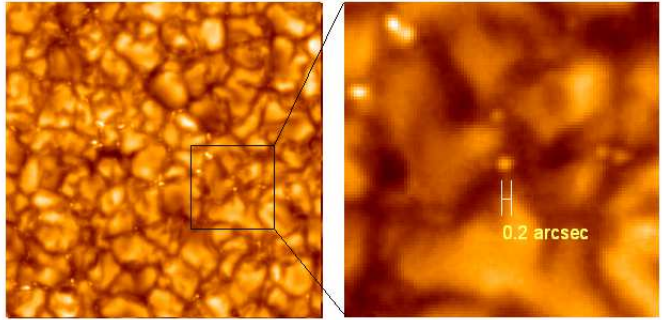}
  \caption[]{\label{ichimoto-fig:gbp}  
G-band (430nm) image obtained by the Broad-band Filter Imager.
}\end{figure}

\begin{figure}  
  \centering
  \includegraphics[width=8cm]{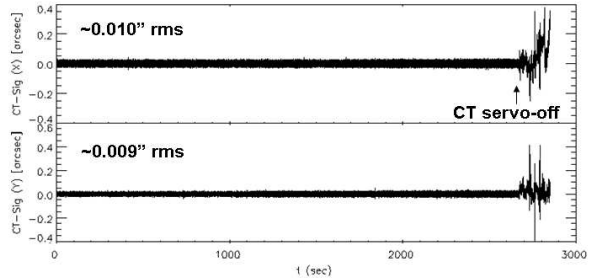}
  \caption[]{\label{ichimoto-fig:xrt}  
Pointing error signals from the correlation tracker (CT) in east-west (top)
and north-south (bottom) direction.
CT servo was turned off around t=2600sec.
}\end{figure}

\subsection{Focus}

\begin{figure}  
  \centering
  \includegraphics[width=10cm]{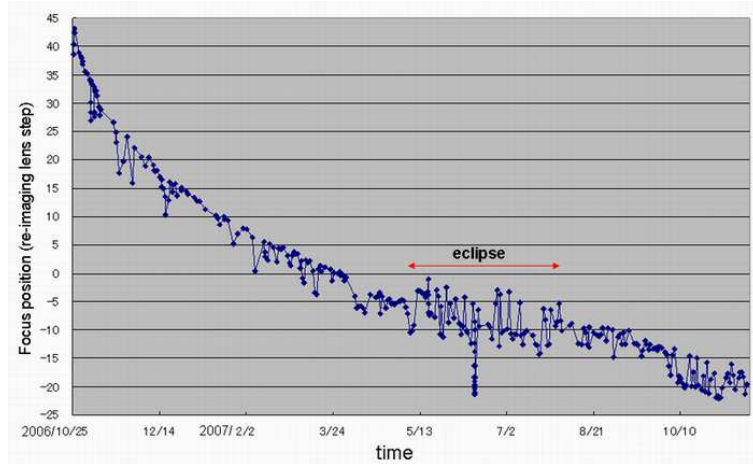}
  \caption[]{\label{ichimoto-focus:xrt}  
Best focus position in G-band 
(reimaging lens position in step) since the first light of SOT.
}\end{figure}

\begin{figure}  
  \centering
  \includegraphics[width=9cm]{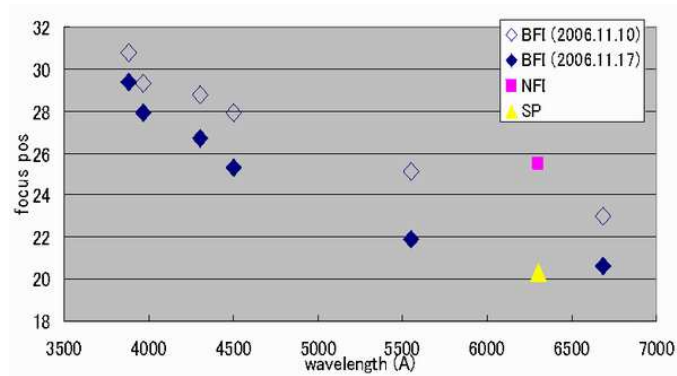}
  \caption[]{\label{ichimoto-fig:chromatic}  
Best focus position as a function of the wavelength.
All wavelengths of BFI and 630.2nm of NFI and SP are plotted.
}\end{figure}

Figure 3 shows the history of the best focus position of the reimaging lens
(step number) in G-band since the first light of the SOT.
Gradual drift of the focus is caused by a dimensional change of the CFRP structure 
of the OTA due to dehydration in space. 
Short term variations are the consequence of change of target region on the sun;
a focus offset by about 5 steps between disk-center and limb pointing is observed.
Though the mechanism for this behavior is not well understood,
the response of the focus change is fast enough to allow us to 
re-adjust the reimaging lens position at each maneuver of satellite in the operation.
During the eclipse season (from early May to early August),
a large drift of focus ($\sim$12 steps) occurs during a day in each orbit.
This is a predicted behavior caused by excursion of the temperature of 
the heat dump mirror in OTA against the day/night cycle.
The eclipse season is certainly a 'degraded performance period' of the SOT.

Figure 4 shows the best focus positions in each wavelength of BFI, 
NFI/630nm and SP.
BFI has a chromatic aberration which was recognized after the launch.
The focus difference between 388nm and 668nm is about 
9 step (=1.36mm of the reimaging lens displacement).
If we set the reimaging lens at the center of the chromatic aberration, 
the focus offset is about 4 step at the longest or shortest wavelength 
and the corresponding wave front error is 21nm rms or $\sim$1/20$\lambda$ at 430nm.
Thus the impact of the chromatic aberration is small, but not completely
negligible when we observe in two extreme wavelengths simultaneously.
There is no evidence of chromatic aberration in NFI, and SP is well co-focused
with the BFI.

\subsection{Spectro-polarimeter} 

The SP achieved a polarimetric accuracy of about 10$^{-3}$ against
the continuum intensity in the 'normal scan mode', i.e., 
with a spatial sampling of 0.16 arcsec/pixel and 4.8sec integration
at each slit position. 
The first column of the polarimeter response matrix (I to QUV cross talk, 
Ichimoto etal. 2007) was determined using the continuum in real data, 
thus the SP is free from the polarization cross-talk at this level.
In nominal observations, we use the quality factor of Q=75 for the on-board 
JPEG compression for the Stokes-I to reduce the amount of 
telemetry data to $\sim$2bit/pixel.
Due to the pixel-to-pixel variation of the CCD sensitivity, 
the lossy compression causes the error of about 1\% in intensity after the
decompression and the flat field correction.
If we use a higher compression quality factor, this error can be reduced, but
Q=75 may be a reasonable compromise in usual observing.

SP has an orbital drift of the spectral image on the CCD with an amplitude 
of about 6 pixels (p-p) in both directions along and perpendicular
to the slit.
The cause of the image shift is a thermal deformation of the FPP structure
according to the orbital phase.
The drift rate is minimized by optimizing the setting of the FPP structural heaters, 
and is finally corrected by the calibration software. 
Thus SP provides impeccable data, though
there is no intrinsic information of the absolute spectral line position.

\subsection{Throughput}
Soon after the first light of the SOT, we confirmed that the light levels 
in each observing path and wavelengths
are close to the values predicted from the pre-launch testing.
After then, however, we observed a gradual loss of throughput 
in shorter wavelengths of BFI; the throughput in Oct. 2007 (1 year after the first light) 
is about 80\% in 388.3 and 396.8nm, 90\% in 430.2nm, 95\% in blue continuum 
and 97-99\% in green and red continuums.
The cause of the loss is not identified at this point,
i.e., we do not find any particular optical component
in SOT showing a significant rise of temperature.
A baking of the FG-CCD did not recover the throughput.
The loss rate seems slightly declining, but further monitoring is necessary.

About 60\% of throughput was lost in 630.2nm of the NFI in a year.
The cause was identified as a degradation of coating on
the blocking filter due to a long exposure to the UV in sunlight;
in the initial phase, we used the 630.2nm line almost all the time 
for observations and testing of the NFI.
Since the blocking filter for 589nm (NaI D1) is durable against the UV, 
this filter is inserted in the beam always during the idle time of the NFI,
thus the degradation of filters is suppressed in the current operation.
There is no evidence of significant loss of throughput in other wavelengths of NFI
and SP.

\subsection{Tunable filter} 

Images of the NFI contain blemishes which degrade or obscure the image 
over part of the field of view. The artifacts are caused by air bubbles 
in the index matching fluid inside the tunable filter. 
They distort and move when the filter is tuned. For this reason, 
NFI observing is usually done in one spectral line at one or a small 
number of wavelengths for extended periods of time. 
Rapid switching between lines is not allowed. 

To suppress the disturbance of bubbles, we are required to block 
4 tuning elements out of 8.
This situation limits our capability of tuning the filter, 
but some useful schemes are still available.
New software to enable such operation was successfully uploaded in Apr. 2007,
and tuning schemes have been developed and tested which permit tuning 
to different positions in a line profile without disturbing the bubbles. 
Figure 5 shows an example of such schemes, i.e., blue and red sides
of NaI D1 (589.6nm) are taken to obtain a Dopplergram and magnetogram.
An example of observed data using this scheme are shown in figure 6,
in which we see clear signals of the Doppler shift and longitudinal
magnetic field in lower chromosphere where NaI D1 is formed.

We note that, though 4 tuning elements are blocked during each line scan,
the Doppler shift due to the orbital motion of the satellite
is to be compensated using all elements.
In the season when the amplitude of the orbital Doppler shift is large
(the worst case is the eclipse season from the early May to the early August),
a long run of such observation occasionally causes a migration of the bubble
to the center of FOV that significantly degrades the data.
Thus the best season for the NFI observation is found in 
October and March when the orbit of satellite is nearly perpendicular 
to the direction towards the sun.

Flat field correction of NFI images is still a challenge, but progress 
is being made on this.
Magnetograms and Dopplergrams are usually self-correcting since 
they are made from ratios of intensity differences.

\begin{figure}  
  \centering
  \includegraphics[width=11cm]{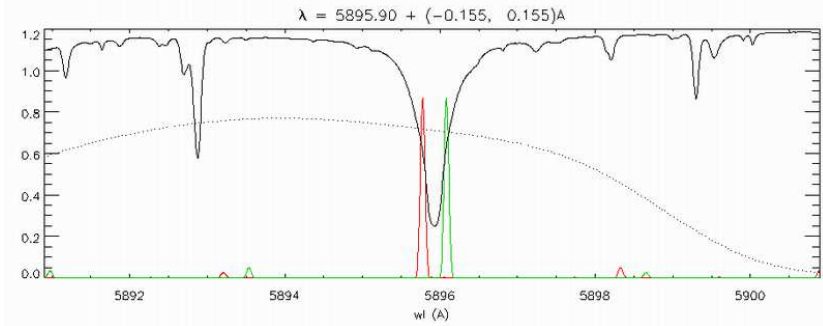}
  \caption[]{\label{ichimoto-fig:tf2}  
An example of tuning scheme for the NFI without disturbing the bubbles.
This is for a simultaneous Dopplergram and magnetogram observation
in blue and red wings of the NaI D1 (589.6nm) line.
}\end{figure}

\begin{figure}  
  \centering
  \includegraphics[width=9cm]{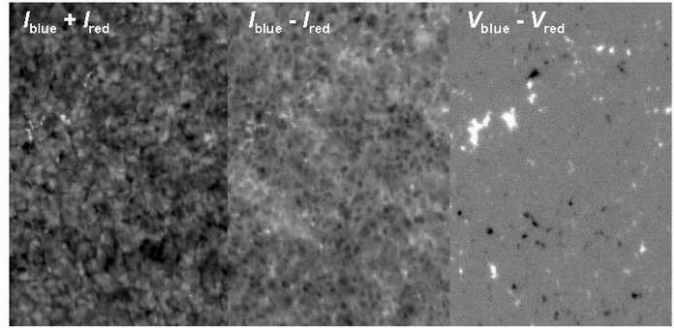}
  \caption[]{\label{ichimoto-fig:NaIVDG}  
Intensity (left), Dopplergram (middle) and magnetogram (right) in NaI 589.6nm 
taken by the NFI with the tuning scheme shown in figure 5.
}\end{figure}

\section{Summary}  \label{ichimoto-sec: summary}

On-orbit performance of SOT was reviewed with some emphasis on 
unexpected features of the instrument. 
In general, 
SOT revealed excellent performance and has been providing outstanding data.
Some unexpected features exist, however; the chromatic aberration in BFI,
degradation of blocking filters in NFI, orbital drift of the SP image 
on the CCD, and air bubbles in the tunable filter.
Most of them are recovered by the operation and dada analysis, 
i.e., frequent focus adjustment, careful heater setting for the FPP structure, 
and corrections in data analysis.
Regarding the NFI, new schemes for tuning the filter without disturbing the bubbles 
have been developed and tested, and some useful procedures to obtain Dopplergram
and magnetogram are now available.
Heavy usage of tunable filter occasionally carries the bubbles into the FOV when the 
Doppler shift by the orbital motion of satellite is large.
The eclipse season (early May to early August) is therefore a degraded performance
period of SOT.
On the other hand, October and March are the best seasons
when the orbit of satellite becomes nearly perpendicular 
to the direction towards the sun.
Monitoring of the throughput and contamination analyses 
of the telescope based on temperatures of optical components in SOT 
will be continued.

\acknowledgements 
Hinode is a Japanese mission developed and launched by ISAS/JAXA,  
with NAOJ as domestic partner and NASA and STFC (UK) as international  
partners. It is operated by these agencies in co-operation with ESA  
and NSC (Norway).






\end{document}